\documentclass[aps,prc,twocolumn,showpacs,showkeys,nofootinbib,superscriptaddress]{revtex4-1}

\usepackage{longtable}
\usepackage{graphicx}
\usepackage{amsmath}
\usepackage{amsfonts}
\usepackage{amssymb}
\usepackage{natbib} 
\usepackage{setspace}
\usepackage{xspace}
\usepackage{cancel}
\usepackage{physics}
\usepackage{tensor,braket,color,bm,slashed}

\usepackage{newtxtext,newtxmath}
\usepackage{mathrsfs}

\usepackage[normalem]{ulem}

\renewcommand{\sout}{\bgroup \color{red} \ULdepth=-.5ex \ULset}

\usepackage{color}

\begin{document}

\title{
 \begin{flushright}
   \rightline{PKNU-NuHaTh-2020-06, LFTC-20-6/58} 
 \end{flushright}
 Effects of neutrino magnetic moment and charge radius constraints and medium modifications of the nucleon form factors on the neutrino mean free path in dense matter }

\author{Parada T.P. Hutauruk}
\email{phutauruk@gmail.com}
\affiliation{Department of Physics, Pukyong National University, Busan 48513, Republic of Korea}
\affiliation{Asia Pacific Center for Theoretical Physics, Pohang, Gyeongbuk 37673, Republic of Korea }
\affiliation{Departemen Fisika, FMIPA, Universitas Indonesia, Kampus UI Depok 16424, Indonesia}

\author{A. Sulaksono}
\email{anto.sulaksono@sci.ui.ac.id}
\affiliation{Departemen Fisika, FMIPA, Universitas Indonesia, Kampus UI Depok 16424, Indonesia}

\author{K. Tsushima}
\email{kazuo.tsushima@gmail.com}
\email{kazuo.tsushima@cruzeirodosul.edu.br}
\affiliation{Universidade Cidade de Sao Paulo (UNICID) and Universidade Cruzeiro do Sul,
01506-000, S\~ao Paulo, SP, Brazil}

\date{\today}

\begin{abstract}
Effects of neutrino charge radius and magnetic moment constraints obtained from the astrophysical observations and reactor experiments, as well as in-medium modifications of the weak and electromagnetic nucleon form factors of the matter on the neutrino electroweak interaction with dense matter, are estimated. We use a relativistic mean-field model for the in-medium effective nucleon mass and quark-meson coupling model for nucleon form factors. We analyze the neutrino scattering cross section and mean free path in cold nuclear matter. We find that the increase of the cross section relative to that without neutrino form factors results in the decrease of the neutrino mean free path when neutrino form factors and the in-medium modifications of the nucleon weak and electromagnetic form factors are simultaneously considered. The quenching of the neutrino mean free path is evaluated to be about 12-58\% for the values of $\mu_\nu = 3 \times 10^{-12} \mu_B$ and $R_\nu = 3.5 \times 10^{-5}~\textrm{MeV}^{-1}$ compared with that obtained for the $\mu_\nu =0$ and $R_\nu =0$. The decrease of the neutrino mean free path is expected to decelerate the cooling of neutron stars. Each contribution of the neutrino form factors to the neutrino mean free path is discussed.
\end{abstract}


\maketitle

\section{Introduction}
\label{intro}
%
In the standard model (SM), neutrinos are assumed to be massless and to have zero magnetic moments and electric charges. However, there are several pieces of evidence for massive neutrinos with nonzero magnetic moments that bound the electron neutrino magnetic moment ($\mu_{\nu_{e}}$) to be smaller than $1.0 \times 10^{\rm -10} \mu_B$ at 90\% confidence level, where $\mu_B$ is the Bohr magneton~\cite{Daraktchieva03}. Note that the suffixes indicating the neutrino weak (flavor) eigenstates as $\nu_e$, $\nu_\mu$, and $\nu_\tau$, will be suppressed hereafter, and will be denoted by $\nu$, otherwise stated. The other stringent bound of $\mu_\nu \lesssim 3.0 \times 10^{\rm -12} \mu_B$ is also available from the study of the red giant population in the globular cluster~\cite{Raffelt99}. The most recent experimental constraint comes from the GEMMA (Germanium Experiment for measurement of Magnetic Moment of Antineutrino) experiment with the upper limit of $\mu_\nu$ smaller than $2.9 \times 10^{-11} \mu_B $ at 90\% confidence level~\cite{Beda13}.

Besides the constraints for neutrino magnetic moment (NMM), the square charge radius for electron neutrino has been reported by LAMPF (Los Alamos Meson Physics Facility) experiment with $R_\nu^2 = (0.9\pm 2.7) \times 10^{\rm -32}$ cm$^2 = (22.5 \pm 67.5) \times 10^{\rm -12}$ MeV$^2$~\cite{Allen92}, while the plasmon decay in the globular cluster star leads to a limit $e_v \lesssim  2 \times 10^{\rm -14} e$, where $e$ is the electron charge~\cite{Raffelt99}. Other recent experimental constraints come from the solar~\cite{Friedland05,Cisneros70,DHH68}, atmospheric~\cite{Fukuda98,BGS89,Hirata88,Aartsen20,Brunner13}, long baseline accelerator and reactor neutrinos of neutrino flavor oscillations~\cite{Ahn02,Yasuda20,Abe14,Acciarri15}, and Supernova SN 1987a Observations~\cite{GAAN87,LC88,BM88}. They clearly show that neutrinos are massive and can have finite magnetic moments and charge radii and hence, the SM has to be extended to accommodate the neutrino mass generation.

Many theoretical models for neutrino mass generation have been proposed over the years, for example, $L-R$ gauge symmetry model~\cite{PS73,BKY87,MM80,MP74,SM75,Mohapatra74,BM89}, flavor dependent gauge symmetry~\cite{HNOO19}, seesaw-type models~\cite{DNN08}, supersymmetry with R-parity breaking~\cite{Barbier04,Martin97,BGH89}, and effective field theory (EFT)~\cite{BT17,LMS20,CT20}. However, the physical origin of neutrino mass generation and neutrino properties have not yet been fully understood. Therefore, more experiments and extensive theoretical studies on neutrino properties and interactions are urgently needed.

Besides the current issues on the neutrino mass generation, several interesting ideas and scenarios on new physics and nonstandard interactions (NSIs) beyond the standard model (BSM) have also been extensively discussed in the literature, namely, neutrino charge radii and magnetic moments~\cite{FS80,KHST05,SWHM06,HYT18,HOT19,Sinev2020,JBG04}, nonstandard interactions~\cite{Sinev2020,ATZ18,MN15,FNP20,EGMSS18}, and neutrino oscillations~\cite{Yasuda20,CFLMMP13,EGMSS18,Wolfenstein77}. In this work, we concentrate on neutrino interaction with the dense matter by considering the neutrino charge radius and magnetic moment constraints together with the medium modifications of nucleon structure as matter constituents, which is an extension work of Refs.~\cite{HYT18,HOT19}.

In the previous work~\cite{SWHM06}, the weak and electromagnetic (EM) interactions of neutrino with dense matter were discussed without considering the in-medium modifications of nucleon form factors. The authors, including two of us, studied the implications of the neutrino form factors in neutrino-matter scattering cross sections via neutral current~\cite{SWHM06} for a bound nucleon by including the effect of the neutrino magnetic moment and charge radius, respectively. In the study, they also considered neutrino trapping. In the present work, the matter is neutron star. Therefore, the neutrino trapping is not considered. They found that the neutrino EM form factors have important role in the differential cross section (DCRS) of electron neutrino and matter scattering for $\mu_\nu \gtrsim 10^{-10} \mu_B$ and $R_{\nu} \gtrsim 10^{-5}$ MeV${^{-1}}$. Furthermore, the effect of neutrino EM form factors is more pronounced at higher nuclear densities.

In recent works of Refs.~\cite{HYT18,HOT19}, we studied medium modifications of the weak and EM form factors of nucleons on neutrino interaction with matter in the absence of the neutrino form factors. The EM form factors reflect their internal structure. We expect that the nucleons and the substructure of the nucleon to be modified by their surrounding environment. Therefore, it is natural to anticipate that the nucleon's EM and weak properties are modified in the nuclear medium.

The lattice QCD has also attempted to study the in-medium modifications of nucleon for a few body systems in Ref.~\cite{CDDG18}. In addition recent experimental observations in electron-nucleus scattering suggest the in-medium modifications of the nucleon EM form factors, although the analysis is yet model dependent~\cite{JLab-A-00,DBBB00,Strauch03,CMPR09,E03-104-10,E03-104-11}. More reviews and discussions on medium modifications of the nucleon can be found in Refs.~\cite{HYT18,HOT19,SBKMV04,Strauch12,BST11,HMPW16}. Triggered by the results of the studies~\cite{SWHM06,HYT18,HOT19} and experimental evidence, in the present work, we address the significance of both, neutrino form factors---magnetic moment and charge radius, and in-medium modifications of the nucleon weak and EM form factors on the neutrino scattering with the matter.

In this paper we employ a relativistic mean-field (RMF) model to describe the matter. As for describing the in-medium nucleon weak and EM form factors, we adopt the quark-meson coupling (QMC) model~\cite{Guichon88}. The RMF and QMC models have been widely and successfully used in the finite nuclei and nuclear matter. We adopt a linear response approximation for describing the scattering of neutrinos with the proton, neutron, electron, and muon, which are the standard constituents of neutron star matter. We then calculate the cross section to estimate the neutrino mean free path (NMFP) by combining both theoretical approaches. In the present work, we consider the NMM and charge radius, which are absent in our previous studies of Refs.~\cite{HYT18,HOT19}, obtained from the observations and reactor experiments together with the density-dependent nucleon form factor.

We find that the neutrino differential cross sections of the electroweak neutrino scattering with the constituents of matter are insensitive to increasing nuclear density. However, the neutrino cross section is sensitive to the magnetic moment and charge radius of neutrino. It gives a counter effect to the in-medium modifications of the nucleon weak and electromagnetic form factor effects. The effect becomes pronounced for larger magnitude of neutrino magnetic moment and neutrino charge radius.

This paper is organized as follows. In Sec.~\ref{scattering}, we briefly present neutrino differential cross sections for scattering with the constituents of matter in the presence of both neutrino form factors (magnetic moment and charge radius) and in-medium modifications of the nucleon weak and EM form factors. In Sec.~\ref{dcrsmfp}, we present the neutrino differential cross section and mean free path. In Sec.~\ref{numerical}, our numerical results are presented, and their implications are discussed. Section~\ref{summary} is devoted for a summary.

\section{Neutral Current Neutrino Scattering with Matter and the Form Factors}
\label{scattering}

Here we briefly review the differential cross section of neutrino scattering with the matter in the weak and EM interactions. In the calculation, we consider neutrino magnetic moment and charge radius, as well as nucleon form factors. Effective Lagrangian density for the neutrino with the constituents of matter in terms of the current-current interaction has a 
form~\cite{HYT18,HOT19}
\begin{eqnarray}
  \label{eq:model1}
  \mathscr{L}_{\rm int}^{j} &=& \frac{G_F}{\sqrt{2}} \left[ l^\mu_{\rm W} \mathcal{J}_{\mu}^{{\rm W} (j)} \right]
  + \frac{4\, \alpha_{\rm em}^{}}{q^2} \left[ l^\mu_{\rm EM} \mathcal{J}_{\mu}^{{\rm EM} (j)} \right], 
  \nonumber \\
\end{eqnarray}
where the first term of the effective Lagrangian is the weak interaction for lepton $l^\mu_W =\left[ \bar{\nu}_e(k^{\prime})\Gamma_{\rm W}^{\mu} \nu_e (k) \right]$, and for nucleon $\mathcal{J}_{\mu}^{{\rm W} (j)} = \left[ \bar{\psi}_j^{} (p^{\prime}) J_{\mu}^{{\rm W} (j)} \psi_j^{} (p) \right]$, and the second term of the effective Lagrangian is the EM interaction for lepton $ l^\mu_{\rm EM} =  \left[ \bar{\nu}_e(k^{\prime})\Gamma^{\mu}_{\rm EM} \nu_e (k) \right]$, and for nucleon $\mathcal{J}_{\mu}^{{\rm EM} (j)}=\left[ \bar{\psi}_j^{} (p^{\prime}) J_{\mu}^{{\rm EM} (j)} \psi_j^{} (p) \right]$, $G_F \simeq 1.166 \times 10^{-5}~\mbox{GeV}^{-2}$ and $\alpha_{\rm em}^{-1} \simeq 137$~\cite{PDG16} are the Fermi (weak) coupling constant and EM fine structure constant, respectively. Here, $\nu_e (k)$ and $\bar{\nu}_e (k^{\prime})$ are respectively the initial and final state neutrino spinors. The $\psi_j^{} (p)$ and $\bar{\psi}_j^{} (p^{\prime})$ stand for the initial and final spinors of the target fermion $j = (n,~p,~e^{-},~\mu^{-})$, and $q$ is the transferred four-momentum. Here, the Dirac neutrino is considered~\cite{Shrock82,Nieves82}, and we will focus on the electron neutrino. Note that we do not consider the neutrino oscillation in the present work.

\subsection{Neutrino Magnetic Moment and Charge Radius}
%
The neutrino weak interaction vertex is defined by $\Gamma_{\rm W}^{\mu} = \gamma^{\mu} \left( 1 -\gamma^{5} \right)$ in Eq.~(\ref{eq:model1}), whereas the EM interaction vertex of Dirac neutrinos is described by four form factors. Applying the current conservation, it then gives~\cite{Nieves82,VE89,GS08a,BGS12}%
\begin{eqnarray}
  \label{eq:model3}
  \Gamma_{\rm EM}^{\mu} (q^2) &=& (f_1 (q^2) + \left(\frac{m_\nu}{m_e} \right) f_2 (q^2)) 
\gamma^\mu + g_{1}^{} 
(q^2) \gamma^\mu \gamma^5, 
  \nonumber \\ && \mbox{}
  - \left[ f_{2}(q^2) + i g_{2}^{}(q^2) \gamma^5 \right] \frac{P^\mu}{2m_{e}^{}}, 
\end{eqnarray}
where $f_{1} (q^2)$, $g_{1}^{}(q^2)$, $f_{2} (q^2)$, and $g_{2}^{} (q^2)$ are the Dirac, anapole, magnetic, and electric form factors, respectively. The $g_1 (q^2)$ = 0 when neutrino mass is not zero. The $P_\mu = k_\mu + k_\mu^\prime$ with $m_\nu^{}$ ($m_{e}^{}$) being the neutrino (electron) mass. Following Refs.~\cite{VE89,FS03,GS08a,BGS12} in the static limit the derivative of the Dirac and anapole form factors against $q^2$ yield respectively the vector charge radius $\braket{R_V^2} =  6 \left. \frac{d f_{1} (q^2)}{dq^2} \right \vert_{q^2=0}$ and the axial-vector charge radius $\braket{R_A^2}=6 \left. \frac{d g_{1}^{} (q^2)}{dq^2} \right\vert_{q^2 =0}$. In the Breit frame, where $q_0^{}$ = 0, it gives $f_{1} (q^2) \simeq \frac{1}{6} \braket{R_V^2} q^2 = - \frac{1}{6} \braket{R_V^2} \bm{q}^{\, 2}$, and $g_{1}^{} (q^2)\simeq \frac{1}{6} \braket{R_A^2} q^2 = - \frac{1}{6} \braket{R_A^2} \bm{q}^{\, 2}$, where $f_{1}(0)= g_1^{}(0) = 0$ are used. The sum $\braket{R^2} \equiv  \braket{R_V^2} + \braket{R_A^2}$ is commonly defined. Further details about neutrino form factors can be found in Refs.~\cite{GKLLSZ15,Nardi03,CHARM-II-95}.

The form factors of $f_{2}(q^2)$ and $g_{2}^{}(q^2)$ at $q^2 = 0$ identify explicitly the neutrino magnetic moment and the charge parity (CP) violating electric dipole moment as $\mu^{m} _\nu = f_{2}(0) \mu_B^{}$ and $\mu_\nu^{e} = g_{2}^{} (0) \mu_B^{}$, where the neutrino magnetic moment is expressed as $\mu_\nu^2 \equiv \left( \mu_\nu^{m} \right)^2 + \left( \mu_\nu^{e}\right)^2$~\cite{KZAM92}, and $\mu_B^{} = e/2m_e^{}$ is the Bohr magneton. The value of the neutrino magnetic moment is estimated around $10^{-10}\mu_B$ as in Refs.~\cite{HSM06,SHM05b,HWSM04,KTAC90,Raffelt99,Borexino-17}.

\subsection{Medium modifications of nucleon form factors}
%

After defining neutrino form factors, here we present the nucleon electroweak form factors in free-space and medium. We briefly present the free-space nucleon form factors, which the weak and EM vertices of the current operators are respectively given by~\cite{HSM06,KPH94,NPR04}, $J_{\mu}^{{\rm W}} = F_{1}^{{W}}(q^2) \gamma_\mu - G_{A}(q^2) \gamma_\mu \gamma^5 + i F_{2}^{{W}}(q^2) \frac{\sigma_{\mu \nu} q^{\nu}}{2M_{N}} + \frac{G_p(q^2)}{2M_N}q_\mu \gamma^5 + \frac{F_3^{W} (q^2)}{2M_N} q_\mu + i \frac{G_2 (q^2)}{2M_N} \sigma_{\mu \nu} q^\nu \gamma_5 $, and $J_{\mu}^{{\rm EM}} = F_{1}^{{\rm EM}}(q^2) \gamma_\mu + i F_{2}^{{\rm EM}}(q^2) \frac{\sigma_{\mu\nu} q^\nu}{2M_N}$, where $M_N$ is the nucleon mass. The scalar form factor $F_3^W (Q^2)$ in the vector part of weak interaction and the axial tensor form factor $G_2 (q^2)$ in the axial part equal to be zero because of the conservation of vector current (CVC) and if we assume SU(2) (isospin) symmetry, that determined on the mass shell. The induced pseudoscalar form factor of $G_p (q^2)$ will be ignored in the present work. This term contributes very small to the cross section, which is proportional to $\mbox{(lepton mass)}^2/M_N^2$~\cite{GS84}. In this work we consider symmetric nuclear matter and the isospin symmetry breaking contribution is small. However, the second class current (proportional $F_3^W(q^2)$ and $G_2(q^2)$) like longitudinal component of the weak vector current is possible to be non-zero in the electro-weak interaction, even in the limit of exact isotopic symmetry, by considering off the mass shell. A violation of isotopic symmetry that associated with the rising additional term of the order of the quark mass difference between the up an down quarks results the nonconservation of the charge-changing component of the vector part of weak current~\cite{Krivoruchenko:2015rza}. The complete formula for free-space weak form factors $F_{1}^{W}(0)$, $G_A(0)$ and $F_{2}^{W}(0)$ and electromagnetic form factors $F_{1}^{\rm EM}(0)$ and $F_{2}^{\rm EM}(0)$ can be found in Refs.~\cite{SWHM06,HYT18,HOT19}. Note that, in this work, $q^2$ dependence of the form factor is weak and neglected, since the medium is being probed at $q^2 = 0$. In such a small transferred momentum of $q^2$, the quark counting rules can be held. The quark counting rules are usually applied in the extremely high $q^2$, although the quark counting rules are not a kind of rigorous theories that directly derived from QCD.

Here we emphasize that we are interested in the in-medium modified nucleon form factors in neutrino-nucleon scattering. Hence, the free-space form factors $G_A(q^2)$, $F_{2}^{ W}(q^2)$, and $F_{2}^{\rm EM}(q^2)$ are replaced by respectively the in-medium modified form factors $G_A^*(q^2)$, $F_{2}^{ W*}(q^2)$, and $F_{2}^{\rm EM*}(q^2)$. However the weak-vector and EM charges $F_1^{ W*} (0)$ and $F_{1}^{\rm EM*}(0)$ do not change (normalized) from the free space. For the sake of consistency the $F_1^{ W*} (q^2)$ and $F_{1}^{\rm EM*}(q^2)$ are kept. Henceforth, those quantities in nuclear medium are represented with an asterisk.

The nucleon form factors in medium are calculated in the QMC model~\cite{HYT18,HOT19,STT05,LTTWS98b,LTT01,Thomas14,Guichon88,STT05,LTTWS98b,LTT01,Thomas14,TSS04,TST00,HT19,HOT18,HOT19-2,TKS03,CCKSKTM13,KTT98,STT96,CJJTW74,Guichon:1995ue,STT05}. Employing the $G$-parity argument, we use the charged-current weak-interaction vector and axial-vector form factors for free-space nucleons. The form factors $F_1^{ W} (q^2)$ and $F_2^{ W} (q^2)$ are assumed to be the same as those of the EM form factors $F_1^{\rm EM}(q^2)$ and $F_2^{\rm EM}(q^2)$, respectively, based on the isospin invariance. The $F_1^{W, \rm EM} (q^2)$ and $F_2^{W, \rm EM} (q^2)$ are connected to the electric $G_E (q^2)$ and magnetic $G_M (q^2)$ Sachs form factors as $G_E(q^2) = F_1^{W, \rm EM}(q^2) + \frac{q^2}{4 M_N^2} F_2^{W, \rm EM}(q^2)$, and $G_M(q^2) = F_1^{W, \rm EM}(q^2) + F_2^{W, \rm EM}(q^2)$, respectively. Note that the induced pseudoscalar form factor $G_P(q^2)$ is dominated by the pion pole and can be calculated using the partial conservation of the axial current (PCAC) relation~\cite{Thomas84,Miller84}.

In the Breit frame the in-medium nucleon form factors are calculated~\cite{LTT01,LTTWS98} using the improved cloudy bag model (ICBM) in Ref.~\cite{LTW97} and the QMC model defines as $G_{E,M,A}^{\rm *\, QMC} (q^2) =  \left( \frac{M_N^{*}}{E_N^{*}} \right)^2 G_{E,M,A}^{\rm *\, sph}( \tilde{q}^2 )$, where the subscriptions of $E$, $M$, $A$, which are the electric, magnetic, and axial-vector form factors, and $\tilde{q}^2 = (M_N^*/E_N^*)^2 q^2$, respectively. The multiplication factor of $(M_N^*/E_N^*)^2$ originally comes from the Lorentz contraction, which the in-medium nucleon energy is expressed as $E_N^{*} = \sqrt{M_N^{*2}(\sigma) + \bm{q}^{2}/4}$. This is obtained from the static spherical MIT bag quark wave functions of nucleons~\cite{LTW97} using the in-medium inputs calculated by the QMC model. The Peierls-Thouless projection, which accounts the center-of-mass and recoil corrections as well as the Lorentz contraction of the internal quark wave function, are considered in the ICBM.

The ratios of the $G_{E,M,A}^{\rm *\, QMC} / (G_{E,M,A}^{\rm ICBM})_{\rm free}^{}$ are then calculated. From these ratios, the in-medium nucleon form factors are determined using the empirical form factors extracted in free space. The in-medium nucleon form factors $G_{E,M,A}^{*} (q^2)$ is obtained as  
$G_{E,M,A}^{*} (q^2) = 
\left[ \frac{G_{E,M,A}^{\rm* \, QMC} (q^2)}{(G_{E,M,A}^{\rm ICMB})_{\rm free}^{} (q^2)} 
  \right] G_{E,M,A}^{\rm emp} (q^2)$.
Here we do not consider the pion cloud effect in the calculation of the axial-vector form factor. The $q^2$ dependence of the normalized axial-vector form factor is reproduced quite well ~\cite{LTT01} by using the empirical parametrization in free space.

\begin{figure}[t]
  \centering\includegraphics[width=0.95\columnwidth]{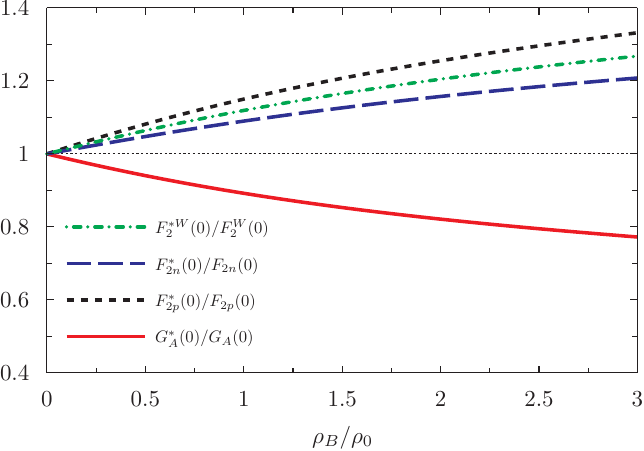}
  \caption{\label{fig1} (Color online.)
    Ratio of the in-medium to free-space nucleon weak and EM form factors at $q^2 = 0$ as a function of $\rho_B^{}/ \rho_0^{}$, where $\rho_0 = 0.15~\mbox{fm}^{-3}$. The figure is adopted from Ref.~\cite{HYT18}.
  }
\end{figure}

In our present work we use the standard values in the QMC model, current quark mass $m_u = m_d = 5~\mbox{MeV}$ (SU(2) isospin symmetry) and the free-space nucleon bag radius $R_N = 0.8~\mbox{fm}$~\cite{Guichon88,STT05,ST94,STT96}. The result for the ratios of the in-medium to free-space nucleon form factors as a function of $\rho_B/\rho_0$ is shown in Fig.~\ref{fig1}. It shows that the form factors of the nucleons ($G_A (0)$, and $F_2 (0)$) are modified in the medium. The ratio of $G_A(0)$ in medium to vacuum decrease as the baryon density increases, whereas the ratios of $F_2 (0)$ in medium to vacuum increase as the baryon density increases. The change of $G_A \simeq \int (\textrm{UU} - \frac{1}{3} \textrm{LL})$ and $F_2 \simeq \int 2 \textrm{UL} $ in medium is due to, in medium, the lower component $\textrm{L}$ of the quark Dirac spinor is enhanced (more relativistic). This leads to decreasing of $G_A$ and increasing of $F_2$, where $\textrm{U}$ is the upper component of the quark Dirac spinor. We remind that the nucleon form factor ratios are obtained using the quark substructure of nucleons. Later the in-medium modified nucleon form factors are used as inputs to calculate the neutrino differential cross sections and mean free path.

\section{Neutrino differential cross section and mean free path}
\label{dcrsmfp}
%

Using the effective Lagrangian density of Eq.~(\ref{eq:model1}), we derive the differential cross section per volume of the neutrino scattering with a target particle. The expression for the differential cross section is given by
\begin{eqnarray}
  \label{eq:model15}
  \left( \frac{1}{V} \frac{d^3 \sigma}{d^2 \Omega^{'} dE_\nu^{'}} \right) 
  &=& - \frac{1}{16 \pi^2} \frac{E_\nu^{'}}{E_\nu} 	\Big[
    \bar{G}_F^2 \left( L_{\nu}^{\alpha\beta} \Pi^{\rm Im}_{\alpha\beta} \right)^{(\rm W)} \nonumber 
\\
    &+& \left( \frac{4\pi \alpha_{\rm em}}{q^2} \right)^2 \left( L_{\nu}^{\alpha\beta} 
    \Pi_{\alpha\beta}^{\rm Im} \right)^{(\rm EM)} \nonumber \\
    &+& \frac{8 \pi \,\bar{G}_F \alpha_{\rm em}}{q^2} \left( L_{\nu}^{\alpha\beta} 
    \Pi_{\alpha\beta}^{\rm Im} \right)^{(\rm INT)} 	\Big], 
  \nonumber \\
\end{eqnarray}
with $E'_\nu$  and $E_{\nu}$ is the neutrino final and initial energies, respectively. The $\left(L_\nu^{\alpha \beta} \Pi_{\alpha \beta}^{\textrm Im} \right)^{(\textrm{W})}$, $\left(L_\nu^{\alpha \beta} \Pi_{\alpha \beta}^{\textrm Im} \right)^{(\textrm{EM})}$, and $\left(L_\nu^{\alpha \beta} \Pi_{\alpha \beta}^{\textrm Im} \right)^{(\textrm{INT})}$ are respectively the contraction of the leptonic and hadronic parts for the weak interaction (W), the contraction of the leptonic and hadronic parts for the EM interaction (EM) and the contraction of the leptonic and hadronic parts for the interference of the weak and EM interactions (INT), where $L_\nu^{\alpha \beta}$ is the leptonic tensor and $\Pi_{\alpha \beta}^{\textrm Im}$ is hadronic polarization tensor. The details of analytic derivations of the polarization tensors and the contraction of the leptonic and hadronic parts for the weak and EM interactions as well as other quantities in Eq.~(\ref{eq:model15}), can be found in Refs.~\cite{SWHM06,SHM05b}.

In the scattering of neutrino with constituents of matter, the nuclear matter is described using the RMF model~\cite{SLVZ03,RPL97,FST96a,GM91,CRD96,Reinhard89,SW86,Serot92,Blaizot80}. The effective mass of the nucleon $M_N^*$ is obtained approximately around 0.78 $M_N$ at the normal nuclear matter density  $\rho_0 = 0.15$ fm$^{-3}$. This value is consistent with those obtained in the QMC model~\cite{STT05,LTTWS98b,LTT01,Thomas14,Guichon88,STT05,LTTWS98b,LTT01,Thomas14,TSS04,TST00,HT19,HOT18,HOT19-2,TKS03,CCKSKTM13,KTT98,STT96,CJJTW74,Guichon:1995ue}. The complete parameter sets used in the RMF model as well as other quantities such as the binding energy, compression modulus and symmetry energy coefficient can be found in Refs.~\cite{SLVZ03,RPL97,FST96a,GM91,CRD96,Reinhard89,SW86,Serot92,Blaizot80}.

The inverse mean free path of the neutrino (INMFP) is easily obtained by integrating the differential cross section of Eq.~(\ref{eq:model15}) over the energy transfer $q_0^{}$ and the three-momentum transfer $\abs{\bm{q}}$. Then, the expression for INMFP at a fixed baryon density is given by~\cite{RPL97}
\begin{align}
  \label{eq:model23}
  \lambda^{-1} (E_\nu) &= \int_{q_0^{}}^{2E_\nu - q_0^{}} d \abs{ \bm {q} } 
  \int_0^{2E_\nu} d q_0^{} \frac{\abs{\bm{q}}}{E'_\nu E_\nu} \frac{2\pi}{V} 
  \frac{d^3 \sigma}{d^2 \Omega' dE'_\nu},
\end{align}
where $E'_\nu = E_\nu - q_0^{}$ is relation between the neutrino final and initial energies. The maximum value of $q_0$ for scattering with nucleons is found to be a function of effective mass and fermi momentum at fixed $\textbf{q}$, it then gives  $q_0^{max} = \sqrt{M_N^{*2} + (p_F + |\textbf{q}|)^2}= E_F \simeq \frac{1}{\sqrt{\left( M_N^{*}/p_F\right)^2 + 1}} |\textbf{q}|$. The maximum value of $q_0$ increases as $M_N^{*}$ decreases at give value of $p_F$. At low density, the fraction of the electrons and muons scattering with neutrinos are very small and the $p_F^{e,\mu}$ are very much larger than their free mass, which implies $q_0^{max} \simeq |\textbf{q}$| and the minimum value of $q_0$ = 0. Hence the range of $q_0$ is larger than that of nucleon targets. This kinematic cutoff is responsible for the sharp peak structure of DCRS of neutrino scattering with the target nucleon. The detailed explanations about the determination of the lower and upper limits in the integral can be found in Ref.~\cite{RPL97}.

\begin{figure*}[t]
  \centering\includegraphics[width=0.97\textwidth]{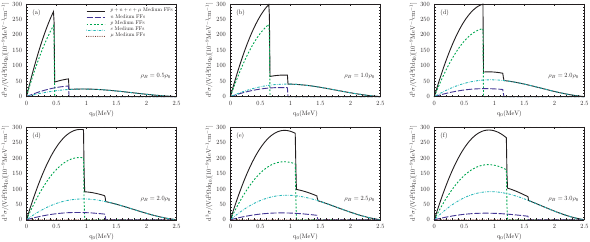}
  \caption{ \label{fig2} (Color online.) Differential cross section of neutrino scattering with constituents of matter for both $\mu_\nu = 5 \times 10^{-10} \mu_B$ and $R_\nu = 3.5 \times 10^{-5} \textrm{MeV}^{-1}$ and in-medium modifications of nucleon form factors as a function of $q_0$ at $|\textbf{q}| = 2.5$ MeV and $E_\nu = 5$ MeV.}
\end{figure*}

\begin{figure*}[t]
  \centering\includegraphics[width=0.97\textwidth]{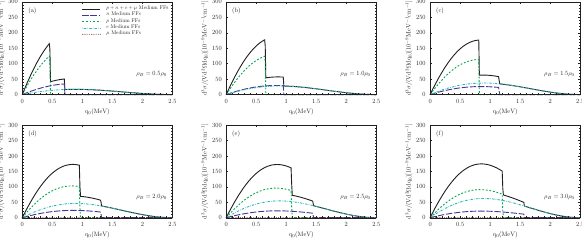}
  \caption{ \label{fig3} (Color online) Differential cross section of neutrino scattering with constituents of matter for both $\mu_\nu = 3 \times 10^{-12} \mu_B$ and $R_\nu = 3.5 \times 10^{-5} \textrm{MeV}^{-1}$ and in-medium modifications of nucleon form factors as a function of $q_0$ at $|\textbf{q}| = 2.5$ MeV and $E_\nu = 5$ MeV.}
\end{figure*}

\begin{figure*}[t]
  \centering\includegraphics[width=0.97\textwidth]{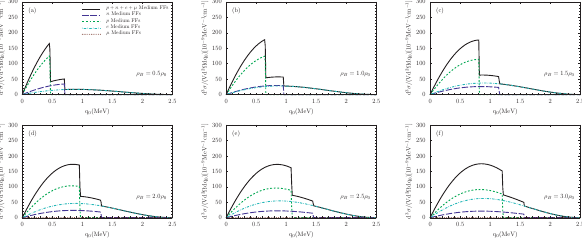}
  \caption{ \label{fig4} (Color online) Differential cross section of neutrino scattering with constituents of matter for both $\mu_\nu = 2.9 \times 10^{-11} \mu_B$ and $R_\nu = 3.5 \times 10^{-5} \textrm{MeV}^{-1}$ and in-medium modifications of nucleon form factors as a function of $q_0$ at $|\textbf{q}| = 2.5$ MeV and $E_\nu = 5$ MeV.
  }
\end{figure*}

\section{Numerical results and discussions \label{numerical}}

We present the differential cross section and mean free path for neutrino scattering with constituents of matter at vanishing temperature by considering neutrino magnetic moment and charge radius and in-medium modified nucleon form factors. Here the cross section and mean free path are evaluated at the energy transfer $q_0^{}$ at $\abs{\bm{q}} = 2.5~\mbox{MeV}$ and the initial neutrino energy $E_\nu = 5~\mbox{MeV}$, which is the typical kinematics for the cooling phase of a neutron star~\cite{HYT18,HP03,NBLM01}. In the later stages of the NS cooling phase, the ground state of NS is reached when the temperature decreases below a few MeV and it becomes cool, dense and highly degenerate. Hence the approximation temperature $T =$ 0 is very good and valid in such a dense system.

In Fig.~\ref{fig2}, we show the total sum of the differential cross sections for $\mu_\nu = 5 \times 10^{-10} \mu_B$ (the same order magnitude with MUNU experiment~\cite{Daraktchieva03}) and $R_\nu = 3.5 \times 10^{-5}~\textrm{MeV}^{-1}$ for different densities. The magnitude of the total sum of the differential cross sections increases dramatically, which is almost twice of the total sum of the differential cross sections for $\mu_\nu = 5 \times 10^{-10} \mu_B$ and $R_\nu =0$ or for $\mu_\nu = 0$ and $R_\nu = 3.5 \times 10^{-5}~\textrm{MeV}^{-1}$. However, the magnitude of the total sum of the differential cross sections does not vary significantly as the nuclear density increases.

In Fig.~\ref{fig3} we show the differential cross section for different densities for $\mu_\nu = 3 \times 10^{-12} \mu_B$ and $R_\nu = 3.5 \times 10^{-5}$ MeV$^{-1}$, which is obtained from the astrophysical bound. Comparing with $\mu_\nu = 0$ and $R_\nu = 3.5 \times 10^{-5}$ MeV$^{-1}$ the total cross section of Fig.~\ref{fig3} does not change drastically. It indicates the effect of the $\mu_\nu = 3 \times 10^{-12} \mu_B$ is less significant for total cross section, as expected. However, the cross 
section for the $\mu_\nu = 3 \times 10^{-12} \mu_B$ and $R_\nu = 3.5 \times 10^{-5}$ MeV$^{-1}$ significantly changes from that with free-space nucleon form factors.

Figure~\ref{fig4} shows the differential cross section of neutrino for $\mu_\nu = 2.9 \times 10^{-11} \mu_B$ and $R_\nu = 3.5 \times 10^{-5}$ MeV$^{-1}$, which are taken from the GEMMA experiment. The shape and magnitude of the cross section are almost the same as in Fig.~\ref{fig3}, as expected.

\begin{figure*}[t]
  \centering\includegraphics[width=1\textwidth]{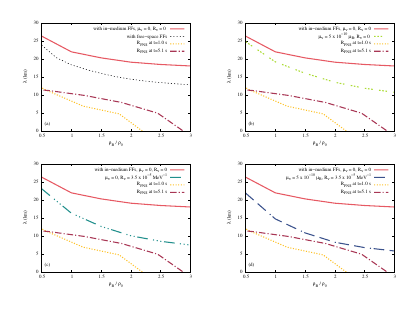}
  \caption{ \label{fig5} (Color online) Neutrino mean free path as a function of nuclear density $\rho_B^{}/ \rho_0^{}$ at $\abs{\bm{q}} = 2.5~\mbox{MeV}$ and $E_\nu = 5~\mbox{MeV}$. The radius of PNS with M$_{\textrm{B}} =$ 1.6 M$_\odot$ for t = 0.1 and 5.1 s is taken from Ref.~\cite{Raduta:2021coc}. }
\end{figure*}

We then observe effects of both neutrino magnetic moment and charge radius and in-medium nucleon weak and EM form factors on NMFP. Figure~\ref{fig5}(a) shows NMFP for the free-space nucleon form factors (dotted line) and in-medium modified nucleon form factors (solid line) with $\mu_\nu = 0$ and $R_\nu = 0$~\cite{HYT18,HOT19} at $\abs{\bm{q}} = 2.5~\mbox{MeV}$ and $E_\nu = 5~\mbox{MeV}$. The result obtained is consistent with the result in Ref.~\cite{HYT18}. As expected, NMFP becomes longer by including the in-medium modified nucleon weak and EM form factors, than that calculated with the free-space nucleon form factors.

The in-medium modifications of the nucleon form factors increase NMFP by 9-38\%~\cite{HYT18,HOT19}. The neutrino-nucleon interaction in a nuclear medium is weaker than those in free space. As we know, the nucleon weak-vector ($F_2^{ W}$) and EM form factors are enhanced in the nuclear medium. However, the quenched axial-vector coupling constant $G_A^*(0)$ dominantly contributes to reducing the total cross section, which results to enhance NMFP.

In Fig.~\ref{fig5}(b) we show NMFP for both $\mu_\nu =0 $ and $R_\nu =0$~\cite{HYT18,HOT19} (solid line) and for $\mu = 5 \times 10^{-10} \mu_B$ and $R_\nu = 0$ (long dashed line) with the in-medium modified nucleon form factors as a function of $\rho_B/\rho_0$. The NMFP decreases as the nuclear density increases. It indicates that the neutrino magnetic moment reduces NMFP by 5-60\%.

Figure~\ref{fig5}(c) shows NFMP decreases as the nuclear density increases for $\mu = 0$ and $R_\nu = 3.5 \times 10^{-5} \mu_B$. This result shows the neutrino charge radius decreases NMFP by 12-58\%.

In Fig.~\ref{fig5}(d) we show NMFP for $\mu_\nu = 5 \times 10^{-10} \mu_B$ and $R_\nu = 3.5 \times 10^{-5}~\textrm{MeV}^{-1}$ as a function of $\rho_B/\rho_0$. The NFMP is more reduced significantly as the nuclear density increases. The neutrino magnetic moment and charge radius reduce NMFP by 15-67\%. This result implies that it decelerates the cooling of neutron stars, meaning that neutrinos need more time to escape from neutron stars.

Comparing with the result of the radius of proto-neutron star (PNS) calculated in the RG(SLy4) model with M$_{\textrm{B}} =$ 1.6 M$_\odot$ for t = 0.1 and 5.1 s~\cite{Raduta:2021coc} and the radius of PNS has less than 20 km in the literature, as in Fig.~\ref{fig5}, our predictions for $\mu_\nu =$ 0 and $R_\nu = $ 0 , $\mu_\nu = 5 \times 10^{-10} \mu_B$ and $R_\nu =$ 0, $\mu_\nu =$ 0 and $R_\nu = 3.5 \times 10^{-5}$ MeV$^{-1}$ , and $\mu_\nu = 5 \times 10^{-10} \mu_B$  and $R_\nu = 3.5 \times 10^{-5}$ MeV$^{-1}$ show that the NMFPs are greater than the PNS radius. However, the PNS becomes less transparent to neutrinos for the $\mu_\nu = 5 \times 10^{-10} \mu_B$  and $R_\nu = 3.5 \times 10^{-5}$ MeV$^{-1}$.

\begin{figure*}[t]
  \centering\includegraphics[width=1\textwidth]{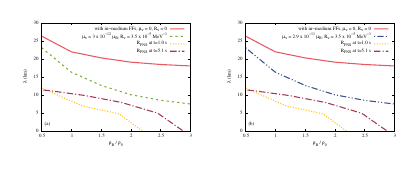}
  \caption{ \label{fig6} (Color online) Neutrino mean free path as a function of nuclear density $\rho_B^{}/ \rho_0^{}$ at $\abs{\bm{q}} = 2.5~\mbox{MeV}$ and $E_\nu = 5~\mbox{MeV}$. The radius of PNS with M$_{\textrm{B}} =$ 1.6 M$_\odot$ for t = 0.1 and 5.1 s is taken from Ref.~\cite{Raduta:2021coc}.
  }
\end{figure*}

In the left panel of Fig.~\ref{fig6} we show NMFP for the $\mu_\nu = 3 \times 10^{-12} \mu_B$ and $R_\nu = 3.5 \times 10^{-5}$ MeV$^{-1}$. The NFMP significantly suppresses with increasing the nuclear density. However comparing with NFMP in Fig.~\ref{fig5}(c), the difference in NMFP is small. It means neutrino charge radius has more significant role than neutrino magnetic moment for NMFP. The contribution with $\mu_\nu = 3 \times 10^{-12} \mu_B$ and $R_\nu =0$ for NMFP is about 1-2\% compared with that for $\mu_\nu =0$ and $R_\nu =0$. Figure~\ref{fig6} shows that the NMFPs for the $\mu_\nu = 2.9 \times 10^{-11} \mu_B$ and $R_\nu = 3.5 \times 10^{-5}$ MeV$^{-1}$ are also greater than the PNS radius, as predicted. Moreover, we observe NMFP for the $\mu_\nu = 2.9 \times 10^{-11} \mu_B$ and $R_\nu = 3.5 \times 10^{-5}$ MeV$^{-1}$ as in the right panel in Fig.~\ref{fig6}. It indicates the contribution of neutrino magnetic moment is less significant for NMFP.

For $\mu_\nu =$ 0 and $R_\nu =$ 0 and free nucleon form factors, the increasing the neutrino energy $E_\nu =$ 50 MeV which is the typical kinematics for the first seconds after the collapse will decrease the DCRS of neutrino. This leads to decrease the NMFP. If we take neutrino energy below 5 MeV i.e. $E_\nu =$ 2.5 MeV, or neutrino energy in the range of a few keV, i.e. $E_\nu \simeq $ [0.2-20] keV, which is typical of the cooling time scale of thousands of years in PNS. Decreasing the neutrino energy will decrease the DCRS of neutrino. It will lead to increase the NMFP. By taking medium nucleon form factor and $\mu_\nu =$ 0 and $R_\nu =$ 0, the DCRS will decrease, which leads the NMFP increases. If we consider in-medium modification nucleon form factor and $\mu_\nu \neq$ 0 and $R_\nu \neq$ 0, it will increase the magnitude of DCRS by factor 3.

\section{Summary \label{summary}}

To summarize, we have studied the effects of magnetic moment and charge radius of neutrino and in-medium modifications of the nucleon weak and electromagnetic form factors on the neutrino scattering differential cross section and mean free path in dense matter. The in-medium-modified nucleon form factors are calculated by the quark-meson coupling model, and the nuclear matter is described in the relativistic mean field model.

The neutrino differential cross sections of the electroweak neutrino scattering with the constituents of matter are insensitive to increasing nuclear density. However, the neutrino cross section is sensitive to the magnetic moment and charge radius of neutrino. It gives a counter effect to the in-medium modifications of the nucleon weak and electromagnetic form factor effects. The effect becomes pronounced for larger magnitude of neutrino magnetic moment and neutrino charge radius. This result is to decrease the neutrino mean free path. The NMFP for $\mu_\nu = 5 \times 10^{-10} \mu_B$ and $R_\nu = 3.5 \times 10^{-5}~\textrm{MeV}^{-1}$ and $\mu_\nu =0$ and $R_\nu =0$ is estimated to be decreased about 15-67\%.

Result for $\mu_\nu = 1 \times 10^{-10} \mu_B$ and $R_\nu=0$, and $\mu_\nu =0$ and $R_\nu = 10^{-5}$ MeV$^{-1}$, which are taken from the reactor experiments of MUNU collaboration, the NMFP are estimated to be reduced 10--15\%  and 3--36\%, respectively, compared to that obtained with $\mu_\nu =0$ and $R_\nu =0$~\cite{HYT18}.

Result for $\mu_\nu = 3 \times 10^{-12} \mu_B$ and $R_\nu=0$, which ($\mu_\nu $) is obtained from the astrophysical observation, the NMFP is shorter than that with the in-medium nucleon form factors. The contribution of the neutrino charge radius is more significant than the neutrino magnetic moment. The contribution of $\mu_\nu =  3 \times 10^{-12} \mu_B$ and $R_\nu = 0$ for NMFP is about 1-2 \% compared to that with $\mu_\nu =0$ and $R_\nu = 0$~\cite{HYT18}.

At higher density the matter compositions in the neutron star cores are expected to change and it will affect the neutrino interaction with constituents of matter. It would be fascinating to include the density dependence of the lepton fractions of matter composition, since the lepton fraction affects the electron neutrino transport in protoneutron stars~\cite{SS13}. These issues will be considered for our future work.

\begin{acknowledgments}
P.T.P.H. was supported by the National Research Foundation of Korea (NRF) grant funded by the Korea government (MSIT) No.~2018R1A5A1025563 and No.~2019R1A2C1005697. The work of A.S. is partly supported by DRPM UI's (PUTI-Q1 and PUTI-Q2) grants No:NKB-1368/UN2.RST/HKP.05.00/2020 and No:NKB-1647/UN2.RST/HKP.05.00/2020. The work of K.T. was supported by the Conselho Nacional de Desenvolvimento Cient\'{i}fico e Tecnol\'{o}gico (CNPq) Process, No.~313063/2018-4, and No.~426150/2018-0, and Funda\c{c}\~{a}o de Amparo \`{a} Pesquisa do Estado de S\~{a}o Paulo (FAPESP) Process, No.~2019/00763-0, and was also part of the projects, Instituto Nacional de Ci\^{e}ncia e Tecnologia -- Nuclear Physics and Applications (INCT-FNA), Brazil, Process. No.~464898/2014-5, and FAPESP Tem\'{a}tico, Brazil, Process, No.~2017/05660-0.
\end{acknowledgments}

\end{document}